\documentclass[aps, prl, twocolumn, titlepage, showpacs]{revtex4}

\usepackage{graphicx}
\usepackage{color}
\usepackage{dcolumn}
\usepackage{amsfonts}
\usepackage{bm}
\usepackage{float}
\usepackage{amsmath}
\usepackage{delarray}

\begin{document}
\title{Superlubric-pinned transition of a two-dimensional solid dusty plasma under a periodic triangular substrate}

\author{Y. Huang$^1$, C. Reichhardt$^2$, C. J. O. Reichhardt$^2$, and Yan Feng$^{1, 3 \ast}$}
\affiliation{
$^1$ Institute of Plasma Physics and Technology, School of Physical Science and Technology, Soochow University, Suzhou 215006, China\\
$^2$ Theoretical Division, Los Alamos National Laboratory, Los Alamos, New Mexico 87545, USA\\
$^3$ National Laboratory of Solid State Microstructures, Nanjing University, Nanjing 210093, China \\
$\ast$ E-mail: fengyan@suda.edu.cn}

\date{\today}

\begin{abstract}

The superlubric-pinned transition in the depinning dynamics of a two-dimensional (2D) solid dusty plasma modulated by 2D triangular periodic substrates is investigated using Langevin dynamical simulations. When the lattice structure of the 2D solid dusty plasma perfectly matches the triangular substrate, two distinctive pinned and moving ordered states are observed, as the external uniform driving force gradually increases from zero. When there is a mismatch between the lattice structure and the triangular substrate, however, on shallow substrates, it is discovered that all of the particles can slide freely on the substrate even when the applied driving force is tiny. This is a typical example of superlubricity, which is caused by the competition between the substrate-particle and particle-particle interactions. If the substrate depth increases further, as the driving force increases from zero, there are three dynamical states consisting of the pinned, the disordered plastic flow, and the moving ordered. In an underdense system, where there are fewer particles than potential well minima, it is found that the occurrence of the three different dynamical states is controlled by the depth of the substrate, which is quantitatively characterized using the average mobility.

\end{abstract}

\maketitle
\section{\uppercase\expandafter{\romannumeral1}. Introduction}
Assemblies of interacting collective particles modified by substrates have been widely studied over the past decades in various two-dimensional (2D) systems, including vortices in type-\uppercase\expandafter{\romannumeral2} superconductors~\cite{harada:1996}, colloidal monolayers~\cite{Reichhardt:2005}, pattern-forming systems~\cite{Reichardt:2003, sengupta:2003}, electron crystals on a liquid helium surface~\cite{monceau:2012}, and dusty plasmas~\cite{Li:2018}. For these physical systems modified by substrates, a variety of new physical phenomena are discovered, such as directional locking~\cite{Reichardt:2004}, superlubricity or the Aubry transition~\cite{Mandelli:2015}, Shapiro steps~\cite{tekic:2010}, anomalous transport~\cite{shaina:2017}, and pinning/depinning dynamics~\cite{Rechardt:2017}. In these studies, the external substrates have various forms, including one-dimensional (1D) periodic substrates~\cite{Reichardt:2015}, 2D periodic substrates~\cite{bechinger:2001}, quasicrystalline substrates~\cite{su:2017}, quasiperiodic substrates~\cite{bohlein:2012}, and even random substrates~\cite{Pertsinidis:2008}. 

In the field of nanoscience, Aubry's theoretical concept~\cite{aubry:1978} for achieving frictionless sliding
is one of the most challenging topics in nanotribology~\cite{persson:1998, vanossi:2013, urbakh:2010}. It is known that the 1D Frenkel-Kontorova model~\cite{braun:1998} consisting of a chain of interacting particles under a static sinusoidal potential exhibits a remarkable dynamical phase transition, first described by Aubry~\cite{aubry:1978}. When the shallowness of the substrate is below a critical value that depends on the precise parameters and incommensurability~\cite{vanossi:2012}, the 1D incommensurate chain-substrate interface can no longer be pinned, indicating that the static friction is zero. The absence of static friction is termed superlubricity~\cite{Peyrard:1983, Dien:2004, fili:2008, Pierno:2015}. When the substrate depth is greater than the critical value, however, the static friction of the studied system is no longer zero, leading to the appearance of a superlubric-pinned transition~\cite{Mandelli:2015}, also called the Aubry transition. Experimentally, the Aubry transition has been observed in various 1D~\cite{Bylinskii:2015, meyer:2015, Bylinskii:2016} and 2D~\cite{nigu:2014, secchi:2016, hirano:1997} systems.

Dusty plasma~\cite{thomas:1996, juan:1996, melzer:1996, fortov:2005, piel:2010, morfill:2009, bonitz:2010, merlino:2004}, also called complex plasma, typically refers to a collection of highly charged micron-sized particles of solid matter in a partially ionized gas. Under laboratory conditions, these dust particles typically are charged to a high negative charge of $\approx - 10^{-4} e$ by absorbing free electrons and ions in plasmas~\cite{feng:2011, qiao:2014}. Their mutual interaction can be described with the Yukawa repulsion~\cite{kono:2000}, also called the Debye-H\"uckel potential, where the shielding effect comes from the free electrons and ions in plasmas. Due to their high negative charges, these dust particles are confined by the electric field of the plasma sheath and can be self-organized into a single layer~\cite{feng:2011, qiao:2014}, forming a so-called 2D dusty plasma. In experiments, these negatively charged dust particles are strongly coupled and exhibit typical solid-like~\cite{feng:2008, hartmann:2014} or liquid-like~\cite{thomas:2004, feng:2010} properties. As the dust particles move inside the plasma gas environment, they always experience a weak frictional gas damping force~\cite{Liu:2003}. Individual particle tracking capabilities have made it possible to study a variety of fundamental physics phenomena using dusty plasmas~\cite{melzer:2001, Hartmann:2010, tsai:2016, wong:2018, he:2020, Hartmann:2013}.

Recently, the collective behaviors of 2D dusty plasmas modified by various periodic substrates have been studied using Langevin dynamical simulations~\cite{Li:2018, wang:2018, Li:2020, Li:2019, Gu:2020, Huang:2022, zhu:2022}. When the interparticle interaction of these dust particles competes with a 1D periodic substrate, a variety of interesting new phenomena are generated, such as splitting of the phonon spectra~\cite{Li:2018}, a structure transition~\cite{wang:2018}, and oscillation-like diffusion~\cite{Li:2020}. If a gradually increasing external driving force is applied to these 1D substrate modulated dust particles, three distinctive dynamical states clearly appear~\cite{Li:2019}, which are the pinned, disordered plastic flow, and moving ordered states~\cite{Gu:2020}. In addition, for a 2D dusty plasma modulated by 2D periodic substrates, various distinctive behaviors caused by the relative motion of particles in each potential well~\cite{Huang:2022} and a direction locking effect~\cite{zhu:2022} are also studied.

A natural next question is whether the Aubry transition also exists in dusty plasmas. The previous Aubry transitions were mainly studied in overdamped colloidal systems~\cite{Davi:2017, Hasnain:2015, Brazda:2018}; however, under underdamped conditions such as those found in 2D dusty plasmas, it is still not clear whether the properties of the Aubry transition or the superlubric-pinned transition~\cite{Mandelli:2015} might be modified. Without an investigation in an underdamped system, the nature of the superlubric-pinned transition cannot be fully understood. Thus, we study the superlubric-pinned transition of a 2D dusty plasma under 2D periodic triangular substrates using various structural and dynamical diagnostics.

In this paper, we report the superlubric-pinned transition of a two-dimensional solid dusty plasma under a periodic triangular substrate using Langevin simulations. In Sec.~\uppercase\expandafter{\romannumeral2}, we briefly introduce our Langevin simulation method. In Sec.~\uppercase\expandafter{\romannumeral3}, we present the obtained results of the superlubric-pinned transition, mainly from various structural and dynamical diagnostics, including the collective drift velocity $V_x$, the 2D distribution function $g(x,y)$, the fraction of sixfold coordinated particles $P_6$, the averaged mobility $\mu$, and the total potential energy per particle $E_{particle}$. Finally, we briefly give our summary of findings in Sec.~\uppercase\expandafter{\romannumeral4}.

\section{\uppercase\expandafter{\romannumeral2}. Simulation method}
Traditionally~\cite{thomas:1996, juan:1996, melzer:1996, fortov:2005, piel:2010, bonitz:2010, merlino:2004, morfill:2009}, 2D dusty plasmas can be characterized using two dimensionless parameters~\cite{San:2001,Oh:2000}, which are the coupling parameter $\Gamma= Q^2/(4 \pi \epsilon_0 a k_B T)$ and the screening parameter $\kappa= a / \lambda_D$. Here, $T$ is the averaged kinetic temperature for dust particles, $Q$ is the charge of one single particle, $a=(\pi n)^{-1/2}$ is the Wigner-Seitz radius~\cite{kal:2004} with the 2D areal number $n$, and $\lambda_D$ is the Debye screening length. To normalize the length, we use either the Wigner-Seitz radius $a$ or the average distance between two nearest neighbors, called the lattice constant $b$.
For the 2D triangular lattice we study here, $b =1.9046a$.

Langevin dynamical simulations are performed to investigate the dynamics of a single layer solid dusty plasma on 2D periodic triangular substrates. In our simulations, for each particle $i$, the equation of motion~\cite{Li:2019} is
\begin{equation}\label{equal_1}
{	m \ddot{\bf r}_i = -\nabla \Sigma \phi_{ij} - \nu m\dot{\bf r}_i + \xi_i(t)+{\bf F}^{S}_i+{\bf F}_d.}
\end{equation}
Here, the particle-particle interaction $-\nabla \Sigma \phi_{ij}$ comes from the binary Yukawa repulsion~\cite{Liu:2003} with $\phi_{ij}=Q^{2} \exp (-r_{i j} / \lambda_{D})/ 4 \pi \epsilon_{0} r_{i j}$, where $r_{ij}$ is the distance between two dust particles $i$ and $j$. The terms $- \nu m\dot{\bf r}_i$ and $\xi_i(t)$ represent the frictional gas drag and the Langevin random kicks ~\cite{Gun:1982, Feng:2008}, respectively. In our simulations, we assume a periodic triangular substrate~\cite{Davi:2017}, which has the form of $W(x, y) = -\frac{2}{9} U_0 [\frac{3}{2}+2 \cos(\frac{2\pi x}{w}) \cos(\frac{2\pi y}{\sqrt3 w}) +\cos(\frac{4\pi y}{\sqrt3 w})]$, where $U_0$ and $w$ correspond to the depth and width of the potential wells, in units of $E_0=Q^{2}/4 \pi \epsilon_{0} a $ and $a$, respectively. As a result, the force from the periodic triangular substrate is just ${\bf F}_i^s = -\frac {8\pi U_0} {9w} \sin (\frac{2\pi x}{w}) \cos (\frac{2\pi y}{\sqrt3 w}) \hat{\bf x} - \frac {8\pi U_0} {9\sqrt3 w} [\cos (\frac{2\pi x}{w}) \sin (\frac{2\pi y}{\sqrt3 w}) + \sin (\frac{4\pi y}{\sqrt3 w})] \hat{\bf y} $, in units of $F_0=Q^{2}/4 \pi \epsilon_{0} a^2 $. The last term on the right-hand side of Eq.~(\ref{equal_1}) is just the external driving force ${\bf F}_d = F_d \hat{\bf x}$, in units of $F_0$. Note, to mimic the dynamics of a single layer solid dusty plasma, in our simulations, all these forces, as well as the particle motion, are completely constrained in a 2D plane.

Our simulation parameters are listed as follows. We specify $N_p = 1024$ particles constrained in a $61.1a \times 52.9a$ rectangular box with the periodic boundary conditions. To reduce the temperature effect on the depinning behavior, the conditions of the 2D dusty plasma are fixed as $\Gamma = 1000$ and $\kappa = 2$, corresponding to a typical 2D Yukawa solid~\cite{ha:2005}. The frictional drag coefficient is fixed to $\nu / {\omega}_{pd} = 0.027$, close to the typical experimental value~\cite{feng:2011}, where ${\omega}_{pd} = (Q^2/2\pi\varepsilon_0 m a^3)^{1/2}$ is the nominal dusty plasma frequency~\cite{kal:2004}. For each simulation run, we integrate $\geq 10^7$ steps with a time step of $0.005{\omega}_{pd}^{-1}$ to obtain the positions and velocities of all particles. 

To quantify the lattice mismatch between the particle number and the 2D substrate, we follow Ref.~\cite{vanossi:2012} and define the mismatch ratio $\rho = w/b$. Due to the periodic boundary conditions, the substrate width $w$ is chosen so that there are integer numbers of potential wells within the simulation box.
For comparison, we focus on three specified mismatch ratio values~\cite{vanossi:2012}, corresponding to the underdense regime with $\rho = 0.89$, the ideally dense regime with $\rho = 1.0$, and the overdense regime with $\rho = 1.1$. In these regimes, the particle number $N_p$ is smaller than, exactly the same as, and larger than the potential number $N_w$, respectively. For our simulations with various values of the substrate depth $U_0$ and the mismatch ratio $\rho$, we gradually increase the external driving force $F_d$ along the $x$ direction from zero. After the simulation system reaches the steady state, we record the particle positions and velocities to calculate various diagnostics of the 2D distribution function $g(x,y)$, the collective drift velocity $V_x$, the fraction of sixfold coordinated particles $P_6$, the averaged mobility $\mu$, and the total potential energy per particle $E_{\rm particle}$. Note that in addition to the results of $N_p = 1024$ reported here, we have also performed a few test runs with $N_p = 4096$ to confirm that our reported results are system size independent.

\section{\uppercase\expandafter{\romannumeral3}. Results and discussion}

\subsection{ {\bf A}. Superlubricity and Aubry transition}

\begin{figure}[htb]
    \centering
    \includegraphics{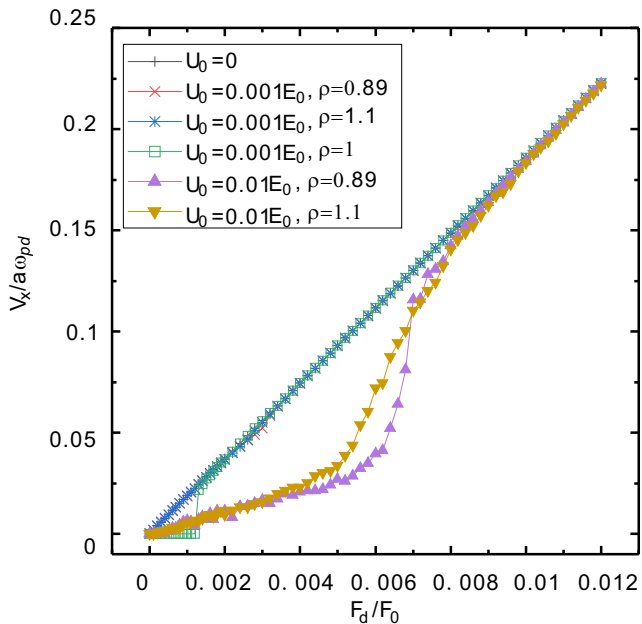}
    \caption{\label{fig:1}
Variation of the collective drift velocity $V_x $ for a 2D Yukawa solid modulated by triangular substrates with various depths $U_0$ and mismatch ratios $\rho$, as the external driving force $F_d$ increases from zero.
Here $\rho$ is defined as $\rho = w/b$, the ratio of the distance between the potential wells $w$ to the lattice constant $b$. For $\rho = 1$ and a small substrate depth of $U_0 = 0.001E_0$, the collective drift velocity $V_x$ is nearly zero at small external driving force $F_d$, indicating that all particles are pinned at the bottom of the potential well. However, for the same depth $U_0 = 0.001E_0$, $V_x$ increases linearly with the increasing external driving force $F_d$ when $\rho = 0.89$ and $1.1$, suggesting that the particles slide freely due to the competition between the substrate and the interaction between particles. As the depth of the substrate increases to $U_0=0.01E_0$, $V_x$ is nearly zero when $F_d$ is small, indicating that all particles are in the pinned state. For comparison, the results of $U_0 = 0$ indicate the response for particles sliding freely without a substrate. Note, the conditions of our simulated 2D Yukawa solid are always $\Gamma = 1000$ and $\kappa = 2$. 
    }
\end{figure}

In Fig.~\ref{fig:1}, we calculate the collective drift velocity $V_x $ for all particles of our simulated 2D solid dusty plasma under triangular substrates, for various values of the depth $U_0$ and the mismatch ratio $\rho$, while the driving force ${\bf F}_d$ increases monotonically. Here, we calculate $V_x$ using  $V_x =N_p^{-1} \langle \sum_{i=1}^{N_p} {\bf v}_i \cdot \hat{\bf x} \rangle $ in units of $a \omega_{pd}$, where ${\bf v}_i$ is the velocity of the particle $i$. Clearly, $V_x$ is the drift velocity only along the direction of the driving force $F_d$. Note, for all of our reported results in this paper, the conditions of the 2D Yukawa solid are always unchanged with $\Gamma = 1000$ and $\kappa = 2$, while the conditions of the substrate and the driving force vary.

For our obtained drift velocity $V_x$ at a mismatch ratio of $\rho = 1$ in Fig.~\ref{fig:1}, two distinctive states are observed, similar to those found for the depinning of 2D dusty plasmas under 1D periodic substrates~\cite{Li:2019, Gu:2020}. At $\rho = 1$, the number of particles is exactly the same as number of potential minima, indicating perfect matching, as shown in Fig.~$2$(c) of Ref.~\cite{Huang:2022}. As shown in Fig.~\ref{fig:1}, for $\rho=1$ at the small driving force $F_d$, the collective drift velocity $V_x$ is nearly zero, indicating that the system is in the pinned state. As the driving force $F_d$ increases further to $F_d=0.0012 F_0$, $V_x$ suddenly jumps directly from 0 to a linearly increasing regime for a substrate depth of $U_0 = 0.001E_0$, where $0.0012 F_0$ is termed the depinning threshold~\cite{Li:2019}. The linearly increasing regime of $V_x$ completely overlaps with the drift velocity for the 2D Yukawa solid with zero substrate or $U_0 = 0$, and the fixed slope of the linear increase is just the frictional gas damping $\nu m$~\cite{Li:2019}. This clearly indicates that all particles slide freely, independent of the 2D periodic triangular substrate, agreeing well with the features of the moving ordered state. We confirm that, when the substrate depth increases for the perfect matching condition of $\rho = 1$, these two states always exist, while the depinning threshold increases monotonically.

Interestingly, in Fig.~\ref{fig:1}, we find that superlubricity occurs~\cite{Peyrard:1983, Dien:2004, fili:2008, Pierno:2015} for mismatch ratios of $\rho = 0.89$ and $1.1$ in our simulated solid dusty plasma under a periodic triangular substrate, where all particles slide freely under the substrate. Clearly, when $\rho = 0.89$ or $1.1$, the particle number is mismatched with the substrate structure. If the substrate depth is small, such as $U_0 = 0.001E_0$, at mismatch ratios of $\rho = 0.89$ or $1.1$ the drift velocity $V_x$ always increases linearly with increasing external driving force $F_d$, suggesting that the particles slide freely and that there is no depinning threshold. In fact, this behavior of the drift velocity is almost identical to the $V_x$ curve for the zero substrate system $U_0 = 0$ in Fig.~\ref{fig:1}. The loss of the depinning threshold $F_{\rm crit}$ for $\rho = 0.89$ and $1.1$ at $U_0 = 0.001E_0$ reflects a typical property of superlubricity, namely, the ability of the particles to slide under any finite driving force $F_d$ on a nonzero substrate due to the competition between the substrate-particle and particle-particle interactions.

If the substrate depth $U_0$ increases further, as shown in Fig.~\ref{fig:1} for $U_0 = 0.01E_0$ at mismatch ratios of $\rho = 0.89$ and $1.1$, the previously observed superlubricity disappears. Here, when $F_d$ is small, $V_x$ is nearly zero, indicating that all particles are in the pinned state, as further confirmed by other diagnostics later. As the driving force $F_d$ increases, two different dynamical states are observed, which are the disordered plastic flow and the moving ordered states, similar to those in Fig.~$2$ of Ref.~\cite{Li:2019}. 

In Fig.~\ref{fig:1}, for $\rho = 0.89$ or $1.1$, when the substrate depth increases from $U_0 = 0.001E_0$ to $0.01E_0$, we find a superlubric-pinned transition, or the Aubry transition~\cite{aubry:1978}. In fact, from the previous investigation in colloids~\cite{vanossi:2012} and our results in Fig.~\ref{fig:1}, it is qualitatively expected that any physical systems under 2D periodic triangular substrates with either overdense ($\rho \textgreater 1 $) or underdense ($\rho \textless 1$) conditions undergo a similar superlubric-pinned transition as a function of increasing substrate depth $U_0$. In the latter sections, we mainly focus on the physics of the underdense condition of our system.

\subsection{{\bf B}.  Three dynamical states}

\begin{figure}[htb]
    \centering
    \includegraphics{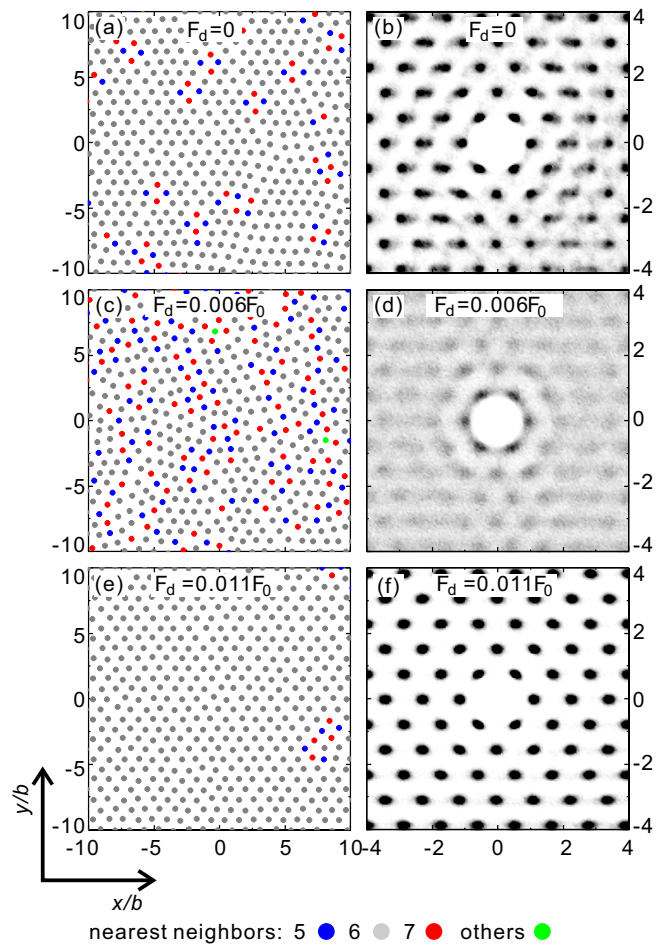}
    \caption{\label{fig:2}
Particle arrangement (a, c, e) and the corresponding 2D distribution functions $g(x,y)$ (b, d, f)  for our simulated 2D Yukawa solid under a triangular substrate with $U_0 = 0.01E_0$ and $\rho =0.89$ driven by different levels of the external force $F_d$. When $F_d = 0$ in (a, b), the particles form an ordered triangular lattice, agreeing well with the pinned state. When $F_d = 0.006F_0$ in (c, d), the particles form a disordered structure, consistent with the disordered plastic flow state. When $F_d = 0.011F_0$ in (e, f), all particles are arranged in a nearly perfect triangular lattice, independent of the locations of the potential wells of the substrate, corresponding to the moving ordered state.
    }
\end{figure}

To study the structure of our simulated dusty plasma solid under 2D periodic triangular substrates in Fig.~\ref{fig:1} for different driving forces, we plot the particle arrangement using their positions and then calculate the corresponding 2D distribution function $g(x,y)$ for our simulated 2D Yukawa solid with the substrate conditions of $U_0 = 0.01E_0$ and $\rho =0.89$, as shown in Fig.~\ref{fig:2}. Here, the 2D distribution function~\cite{loudiyi:1992} $g(x,y)$ is the static structural measure widely used for anisotropic systems such as the system studied here, and it provides the probability density of finding a particle at a 2D position relative to a chosen central particle. Through comparison with the drift velocity results in Fig.~\ref{fig:1}, there are clearly three typical dynamical states consisting of the pinned, the disordered plastic flow, and the moving ordered that appear during the depinning of a 2D solid dusty plasma under a 2D periodic triangular substrate. Figure~\ref{fig:2} confirms these three states directly from the particle arrangements and the corresponding 2D distribution function $g(x,y)$. 

When the external driving force $F_d = 0$, as in Figs.~\ref{fig:2}(a) and~\ref{fig:2}(b), most of the particles have six nearest neighbors, with only a few defects scattered randomly, and the corresponding $g(x,y)$ exhibits a highly ordered structure. These features agree well with the properties of the pinned state. When the driving force is larger, as at $F_d = 0.006 F_0$ in Figs.~\ref{fig:2}(c) and \ref{fig:2}(d), a large number of particles no longer have six neighbors and the corresponding $g(x,y)$ has disordered features, suggesting that some particles escape from the cages formed by their neighbors, leading to the disordered plastic flow state. When the driving force is high enough to overcome the triangular substrate, such as at $F_d = 0.011 F_0$ in Figs.~\ref{fig:2}(e) and \ref{fig:2}(f), almost all particles have six nearest neighbors and the corresponding $g(x,y)$ exhibits a highly ordered structure again, indicating that the system is in the moving ordered state, so that all particles form a nearly perfect triangular lattice, independent of the locations of the potential wells of the substrate. Thus, the three different states inferred from Fig.~\ref{fig:1} are further confirmed by the structure measures in Fig.~\ref{fig:2}. Note, a similar trio of typical dynamical states are also observed in a defective flux-line lattice ~\cite{shi:1991}, Skyrmions~\cite{reichardt:2015}, superconducting vortices~\cite{Thorel:1973}, vortex lattices~\cite{Koshelev:1994}, and the depinning of 2D dusty plasmas on 1D periodic substrates~\cite{Li:2019, Gu:2020}.

\subsection{{\bf C}.  Superlubric-pinned transition}

\begin{figure}[htb]
    \centering
    \includegraphics[width=3.1in]{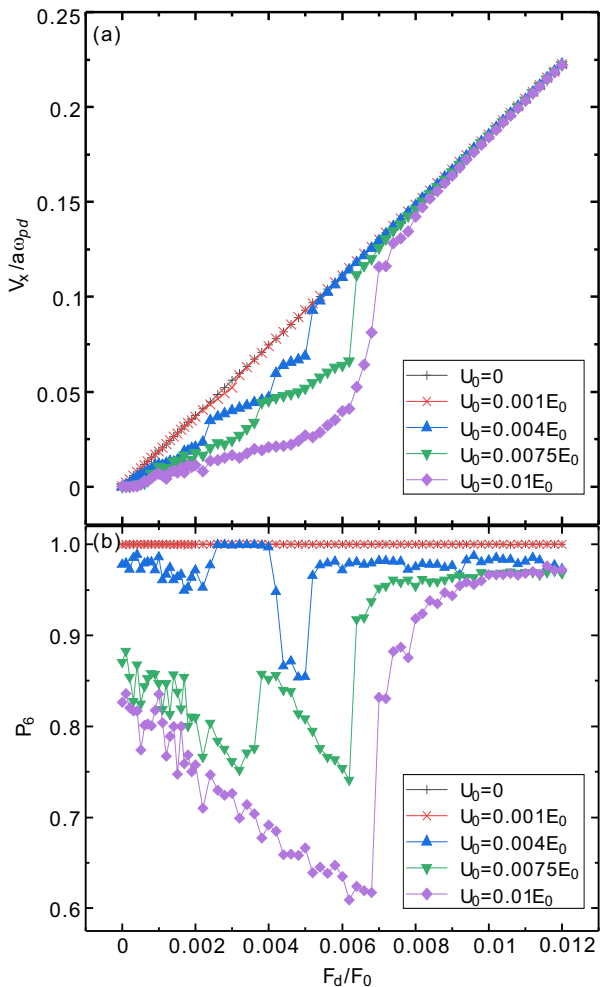}
    \caption{\label{fig:3}
Variation of the collective drift velocity $V_x$ (a) and the corresponding fraction of sixfold coordinated particles $P_{6}$ (b) for the 2D Yukawa solid modulated by triangular substrates with various depths $U_0$ as the external driving force $F_d$ increases for a fixed mismatch ratio $\rho=0.89$. For small substrate depths of $U_0 = 0$ and $U_0=0.001E_0$, the linear increase of $V_x$ with $F_d$ and the unchanging value of $P_{6}=1.0$ both indicate that the system is always in the moving ordered state. When the substrate depth increases to a larger value of $U_0 = 0.004E_0$, at lower driving forces $F_d < 0.005 F_0$ the values of $V_x$ and $P_6$ are depressed below the values found at $U_0 = 0.001E_0$, clearly indicating that disordered plastic flow occurs. However, at higher driving forces $F_d > 0.005 F_0$, $V_x$ and $P_{6}$ rise to match the values found at $U_0 = 0.001E_0$, clearly indicating that the moving ordered state forms. As the depth of the substrate increases to $U_0=0.0075E_0$ or further to $U_0=0.01E_0$, the variations of $V_x$ and $P_{6}$ with $F_d$ clearly indicate that the three states of pinned, disordered plastic flow, and moving ordered lattice all appear.
    }
\end{figure}

To focus on the dynamics of the underdense regime, we fix the mismatch ratio to $\rho = 0.89$ and then calculate the collective drift velocity $V_x $ and the corresponding fraction of sixfold coordinated particles $P_{6}$ as shown in Fig.~\ref{fig:3} for our simulated 2D solid dusty plasma under triangular substrates with various depths $U_0$. Here, $P_{6}$~\cite{Reichhardt:2005} is defined as $P_6 = N_{p}^{-1} \langle \sum_{i=1}^{N_p} \delta (6 - z_{i}) \rangle$, where $z_i$ is the coordination number of particle $i$ obtained from the Voronoi construction. For a perfect 2D triangular lattice, $P_6 = 1$, while the value of $P_6$ is reduced for a more disordered 2D system.

From our obtained drift velocity $V_x$ and the corresponding $P_6$ at mismatch ratio $\rho = 0.89$ in Fig.~\ref{fig:3}, we further confirm the appearance of the three dynamical states described above. For a shallow substrate depth of $U_0 = 0.001 E_0$ in Fig.~\ref{fig:3}, the collective drift velocity $V_x $ always increases linearly with the driving force $F_d$ and the corresponding $P_6$ is always $P_6 \approx 1$, indicating that all of the particles slide freely and the lattice is highly ordered. In fact, the results of $V_x$ and $P_6$ for $\rho = 0.89$ and $U_0 =0.001E_0$ almost exactly match those found for $U_0 = 0$, further suggesting that the system is in the moving ordered state. For a deeper substrate of $U_0 = 0.004 E_0$, the decay of $P_6$ to a reduced value and the relatively steep increase of $V_x$ over the range $F_d \textless 0.005F_0$ in Fig.~\ref{fig:3} suggest that some particles overcome the constraint of the substrate and the cages formed by their neighboring particles, so that disordered plastic flow occurs. However, for large driving forces $F_d \textgreater 0.006 F_0$, $V_x$ increases linearly with $F_d$ and the corresponding $P_6$ goes back to 1 again, clearly indicating that the system reaches the moving ordered state. If the substrate depth increases further to $U_0=0.0075E_0$ or even to $U_0=0.01E_0$, the $V_x$ and $P_6$ curves in Fig.~\ref{fig:3} indicate that all three of the distinctive dynamical states occur. When $F_d$ is small, $V_x$ is nearly zero and $P_6$ is relatively high with $P_6 \textgreater 0.8$, so that the system is in the typical pinned state. When $F_d$ increases to an intermediate level of $F_d \textless 0.009 F_0$, we clearly observe that $V_x$ increases more steeply and the value of $P_6$ decreases substantially, corresponding to the disordered plastic flow state. As the driving force increases further to $F_d \textgreater 0.01 F_0$, $V_x$ increases linearly with $F_d$ and the corresponding $P_6$ goes back to high values close to 1, in good agreement with the moving ordered state.

Based on the results in Fig.~\ref{fig:3}, we find that the occurrence of three dynamical states depends not only on the value of the driving force $F_d$ but also on the depth of the substrate $U_0$, as shown in Fig.~\ref{fig:3}. If the substrate depth is shallow, such as $U_0 = 0.001 E_0$, the pinned state disappears completely, reflecting the typical property of superlubricity, and the moving ordered state always occurs even for the lowest driving force $F_d$. From our interpretation, this superlubricity for the mismatch ratio of $\rho \neq 1$ is attributed to the increased repulsive interaction between particles, which is able to overcome the forces from the substrate on the particles. If the substrate depth increases further to $U_0 = 0.004 E_0$, the constraint from the substrate is enhanced and the particle arrangement is modified to a more disordered structure, resulting in the observed disordered plastic flow state. When the driving force $F_d$ is large enough to completely overcome the constraint from the substrate, the moving ordered state emerges. If the substrate depth increases further to $U_0=0.0075E_0$ or even to $U_0=0.01E_0$, the constraint from the substrate is large enough to strongly confine all particles, leading to the pinned state. As the driving force $F_d$ gradually increases from zero to higher values beyond the depinning threshold, the plastic flow state occurs first, and then the moving ordered state occurs.

\begin{figure}[htb]
    \centering
    \includegraphics{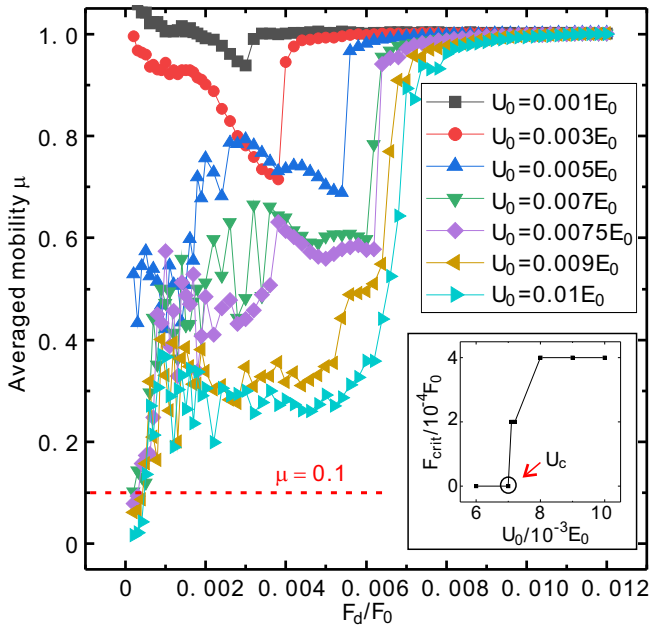}
    \caption{\label{fig:4}
Obtained averaged mobility $\mu$ of our simulated 2D Yukawa solid under triangular substrates with varying substrate depth $U_0$ and fixed mismatch ratio $\rho = 0.89$ for increasing external driving force $F_d$.
Clearly, for the typical shallow substrate with $U_0 = 0.001E_0$, when $F_d$ gradually increases from zero, the averaged mobility $\mu \approx 1$, suggesting that the system is always in the moving ordered state.
As the substrate depth $U_0$ increases, $\mu$ decreases substantially, and a disordered plastic flow occurs.
As the substrate depth $U_0$ further increases beyond the critical Aubry depth of $U_c = 0.007E_0$, the particles are strongly pinned to the substrate while $F_d$ is small, clearly indicating the formation of a pinned state. Due to the random motion of the particles, we follow Ref.~\cite{Brazda:2018} and define the particles to be sliding freely until $\mu$ drops below 10\% of the $\mu=1$ free sliding limit. The dashed line indicates the value of $\mu$, below which the system is defined to be pinned. The inset in the lower right corner shows the depinning threshold $F_{\rm crit}$ as the substrate depth $U_0$ increases from zero.
    }
\end{figure}

As presented in Fig.~\ref{fig:4}, to better define the pinned state of our simulated system, we calculate the averaged mobility $\mu$ of our simulated system for the fixed mismatch ratio $\rho = 0.89$ at various substrate depths $U_0$, as the external force ${\bf F}_d$ increases gradually. Here, $\mu $~\cite{vanossi:2012} is defined as the ratio of the collective drift velocity $V_x$ to the driving force $F_d$,
\begin{equation}
\label{equal_2}
\begin{aligned}
\mu= \frac {\nu m V_x} {F_d},  
\end{aligned}
\end{equation}
where $\nu $ is the frictional drag coefficient. In the absence of a substrate, the averaged mobility $\mu$ of our simulated system should always be around unity, since the driving force $F_d$ is completely balanced by the frictional gas damping $\nu m V_x$.

The averaged mobility $\mu$ at $\rho=0.89$ in Fig.~\ref{fig:4} indicates that three distinctive states clearly appear. For a shallow substrate depth of $U_0 = 0.001 E_0$ in Fig.~\ref{fig:4}, we find that the averaged mobility $\mu \approx 1$, indicating that the driving force $F_d$ equals the frictional gas damping $\nu m V_x$. This suggests that all particles slide freely under the confinement of the triangular substrate, corresponding to the moving ordered state. If the substrate depth $U_0$ increases further to $U_0=0.005 E_0$, $\mu$ decreases substantially when the driving force $F_d$ is small due to the enhancement of the confinement from the substrate. However, when the driving force increases to $F_d \textgreater 0.006 F_0$, $\mu$ goes back to $\mu \approx 1$, indicating that the system reaches the moving ordered state. If the substrate depth $U_0$ further increases to $U_0=0.01 E_0$, at small $F_d$ there is a substantial decrease in $\mu$ to much lower values very close to 0. Here the value of $V_x$ is nearly zero, corresponding to the pinned state. As $F_d$ increases to $F_d\approx 0.0009 F_0$, the value of $\mu$ increases sharply to a value $\mu\approx 0.3$, and remains in this range even when the driving force increases to $F_d=0.006 F_0$. The nonzero value of $\mu$ that is substantially smaller than unity indicates that the driving force $F_d$ is higher than the frictional gas damping $\nu m V_x$, so that only a portion of particles are able to move in response to the driving force, corresponding to the plastic flow state. When $F_d$ further increases beyond this intermediate range, the value of $\mu$ increases abruptly to a value close to unity, indicating that the system has entered the moving ordered state. Note that in comparison with the mobility for overdamped colloidal systems, such as in Ref.~\cite{vanossi:2012}, although the general trend of the variations in mobility we observe is almost the same, our mobility results for dusty plasmas seem to be much more noisy. We attribute this noisy feature of the mobility to the underdamping of the particle motion, since the fluctuations of the particle velocity is much more substantial than what is found for the overdamped colloids in Ref.~\cite{vanossi:2012}.

Here we follow the criterion suggested in Ref.~\cite{Brazda:2018} to distinguish the plastic flow state from the pinned state using the obtained mobility results in Fig.~\ref{fig:4}. In Ref.~\cite{Brazda:2018}, it is suggested that a mobility of $\mu \textless 10\%$ means that the particles are pinned. As a result, a criterion of $\mu = 10\%$ can be used to divide the pinned and the disordered plastic flow states, as indicated by the dashed line in Fig.~\ref{fig:4}. The intersection between the obtained mobility results and $\mu = 10\%$ corresponds to the critical driving force $F_{\rm crit}$ where the pinned and the disordered plastic flow states both occur. The inset of Fig.~\ref{fig:4} presents our obtained $F_{\rm crit}$ results for the varying substrate depth $U_0$ when $\rho = 0.89$. Clearly, as the substrate depth increases, our obtained $F_{\rm crit}$ values increases from 0 to higher values. Here, $U_c$ is the critical substrate depth value, often called the critical Aubry depth $U_c$~\cite{vanossi:2012, Brazda:2018}, beyond which the corresponding $F_{\rm crit}$ is higher than 0, i.e., the pinned state starts to occur when the substrate depth $U_0 \textgreater U_c$. However, if $U_0 \textless U_c$, then $F_{\rm crit}$ is always zero, indicating that the pinned state no longer occurs. That is, the substrate is not able to confine particles even under a very tiny driving force, and superlubricity occurs.
 
\begin{figure}[htb]
    \centering
    \includegraphics{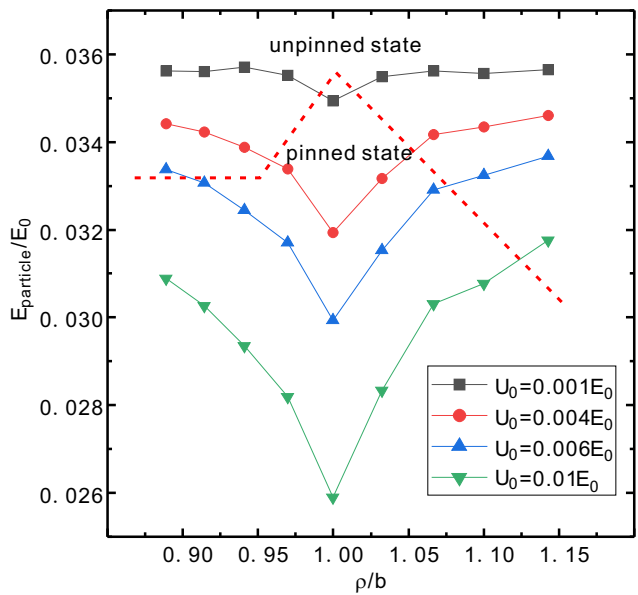}
    \caption{\label{fig:5}
Obtained total potential energy $E_{\rm particle}$ per particle of our simulated 2D Yukawa solid under triangular substrates with varying substrate depth $U_0$ and increasing mismatch ratio $\rho$.
Here, $E_{\rm particle}$ is defined to be the sum of the averaged particle-particle repulsion $U_{pp}$ and the averaged particle-substrate interaction energy $W_{sp}$. Clearly, for each depth $U_0$, when $\rho$ is close to 1, $E_{\rm particle}$ decreases sharply, suggesting that all particles are strongly pinned at the bottom of the substrate because the arrangement of particles is perfectly matched with the substrate configuration. When the substrate depth $U_0$ is small, such as $U_0 = 0.001E_0$, the variation of $E_{\rm particle}$ indicates that the system undergoes a superlubric-pinned-superlubric transition while $\rho$ increases due to the competition between the substrate-particle and particle-particle interactions. As the substrate depth $U_0$ increases gradually, $E_{\rm particle}$ decreases substantially, indicating that the constraint of the substrate on the particles is enhanced, and the pinned state occurs when $\rho \neq 1$. Based on the criterion of the mobility $\mu$, as in Fig.~4, we draw a dashed line to indicate the boundary between the pinned and unpinned states.
    }
\end{figure}

To study the transition of the static structure for our simulated 2D solid dusty plasma under a triangular substrate, we calculate the potential energy per particle $E_{\rm particle}$ in Fig.~\ref{fig:5} while the mismatch ratio $\rho$ and the substrate depth $U_0$ both vary. Here, $E_{\rm particle}$ is the summation of both the averaged particle-particle repulsive potential $U_{pp}$ and the averaged particle-substrate potential $W_{sp}$, both calculated from the obtained particle positions in simulations. In Fig.~\ref{fig:5}, to distinguish between the pinned and unpinned states, we draw a dashed line based on the criterion of the obtained mobility of $\mu = 0.1$, as described above. For all parameters above this dashed line, the pinned state never occurs, so that superlubricity always happens. However, as the depth of the substrate increases to $U_0=0.0075E_0$ or further to $U_0=0.01E_0$, a pinned state appears below this dashed line, leading to the appearance of the plastic flow and the moving ordered states as the driving force increases from zero.

The transition between the pinned state and the unpinned, or superlubric, state for different conditions can be clearly identified from Fig.~\ref{fig:5}. Regardless of the value of the substrate depth $U_0$, when the mismatch ratio $\rho$ is close to unity, $E_{\rm particle}$ decreases sharply, suggesting that the particles are strongly pinned at the bottom of the potential well. If the substrate is shallow, such as for $U_0 = 0.001E_0$, the variation of $E_{\rm particle}$ indicates that as the mismatch ratio $\rho$ gradually increases, the system undergoes a transition from the superlubric to the pinned, and then to the superlubric state again. We attribute this superlubric-pinned-superlubric transition to the competition between the substrate-particle and particle-particle interactions. If the substrate depth $U_0$ increases further to $U_0=0.01E_0$, $E_{\rm particle}$ exhibits a much more pronounced downward trend, suggesting that the confinement from the substrate is greatly enhanced, even when $\rho \neq 1$. In addition, we also find that for a given substrate depth $U_0$, it is more difficult to achieve the pinned state when $\rho \gg 1$ or $\rho \ll 1$, probably due to the extreme mismatch between the lattice structure and the substrate.

\section{\uppercase\expandafter{\romannumeral4}. Summary}
In summary, using Langevin dynamical simulations, we find a superlubric-pinned transition in the depinning dynamics of a 2D solid dusty plasma modulated by 2D triangular periodic substrates while the mismatch ratio varies. For a mismatch ratio of unity, from the calculated overall drift velocity we observe two distinctive states: the pinned and the disordered plastic flow. If the substrate is shallow, however, then for mismatch ratios of $\rho=0.89$ or $\rho=1.1$, the pinned state completely disappears and all particles are able to slide freely on the substrate even when the applied driving force is tiny, consistent with superlubricity. We attribute this superlubricity to the competition between the substrate-particle and particle-particle interactions. If the substrate depth increases further, a gradual increase of the driving force from zero produces three dynamical states of the pinned, the disordered plastic flow, and the moving ordered states.

In the analysis of the dynamics in the underdense regime with a mismatch ratio of $\rho=0.89$, we find that the occurrence of three dynamical states is strongly controlled by the substrate depth. The finding is obtained from various diagnostics, including the 2D distribution function, the collective drift velocity, the fraction of sixfold coordinated particles, the averaged mobility, and the total potential energy per particle. If the substrate depth is shallow, the system is always in the moving ordered state, leading to our observed superlubricity. If the substrate depth increases further, the disordered plastic flow state begins to appear at small driving forces and there is a transition to a moving ordered state at larger driving forces. If the substrate depth increases further, as the driving force increases from zero, three dynamical states are clearly observed. Previous studies of superlubric-pinned transitions focused on overdamped systems; however, our current simulations clearly show that this transition also occurs in underdamped systems. Our simulation results suggest that the superlubric-pinned transition may be realized in future dusty plasma experiments.

\section{Acknowledgments}

The work was supported by the National Natural Science Foundation of China under Grant No. 12175159 and No. 11875199, the 1000 Youth Talents Plan, startup funds from Soochow University, and the Priority Academic Program Development of Jiangsu Higher Education Institutions, and the U. S. Department of Energy through the Los Alamos National Laboratory. Los Alamos National Laboratory is operated by Triad National Security, LLC, for the National Nuclear Security Administration of the U. S. Department of Energy (Contract No. 892333218NCA000001).

\end{document}